%% file: Dusty_Plasma_Revised.tex
\documentclass [aps,onecolumn,showpacs,reprint,pre]{revtex4-1}
\usepackage{graphicx}
\usepackage{textcomp}
\usepackage{psfrag}
\usepackage{float}
\usepackage{amsmath}
\usepackage{mathtools}
\usepackage{xcolor}
\usepackage{url}
\usepackage{hyperref}
\usepackage{caption}
\usepackage[us,12hr]{datetime}

\begin{document}
\title{Phase transition of three-dimensional finite-sized charged dust clusters in a plasma environment}
\author{Hirakjyoti Sarma}
\email{hirakphy2019@gmail.com}
\affiliation{Department of Physics, Tezpur University, Napaam, Tezpur- 784028, India}
\author{Ritupan Sarmah}
\email[]{ritupan@tezu.ernet.in}
\affiliation{Department of Physics, Tezpur University, Napaam, Tezpur- 784028, India}
\author {Nilakshi Das}
\email{ndas@tezu.ernet.in}
\affiliation{Department of Physics, Tezpur University, Napaam, Tezpur- 784028, India}
\date{\today}

\begin{abstract}
	The dynamics of a harmonically trapped three-dimensional Yukawa ball of charged dust particles immersed in plasma  is investigated as function of external magnetic field and Coulomb coupling parameter using molecular dynamics simulation. It is shown that the harmonically trapped dust particles organize themselves into nested spherical shells. The particles start rotating in a coherent order as the magnetic field reaches a critical value corresponding to the coupling parameter of the system of dust particles. The magnetically controlled charged dust cluster of finite size undergoes a first-order phase transition from disordered to ordered phase. At sufficiently high coupling and strong magnetic field, the vibrational mode of this finite-sized charged dust cluster freezes, and the system retains only rotational motion. 
\end{abstract}


\maketitle

\section{Introduction}
It is well known that the dynamics of systems driven far from equilibrium depend on external stimuli. Interesting dynamics may arise in a heterogeneous spatially extended system due to the emergence of collective modes that depend on the nature of instability \cite{cross1993pattern}\cite{ghosh2019emergent}\cite{ananthakrishna2007current}\cite{sarmah2013influence}\cite{sarmah2015correlation}\cite{doi:10.1146/annurev-conmatphys-070909-104101}. Dusty plasma is an ideal platform to study the physics of both extensive and nonextensive systems (particles interacting via noncollective long-range forces), such as the formation of structures, transitions from ordered to disordered states, stability, etc. The physics of dust clusters may be relevant for the development of microstructures, nano-materials, ions in traps, atomic clusters, etc \cite{vladimirov2007non}. Both temporal and spatial scale lengths are stretched in the dusty plasma which can be attributed to the comparatively large size and mass of the dust particles and observation of the phenomena becomes much easier in the laboratory in such a system. Although phase transition is usually studied in the bulk system, it can still be of interest in a finite system, exhibiting novel features and revealing underlying physics. The dust particles immersed in plasma get charged by the flow of plasma ions and electrons or due to radiation in astrophysical systems. The presence of such grains may significantly affect the overall collective behavior of plasma. On the other hand, the plasma particles, specifically the ions, mediate the interaction among dust grains and this may often lead to the formation of structures like crystals, clusters, vortices, etc. Whether the system behaves like a collective or noncollective system depends upon the number of particles present and the geometry of the system. A one-dimensional(1D) string of dust or a two-dimensional dust cluster often behaves as a nonextensive system where dust-dust interaction reflects the properties of ion traps, quantum dots, etc.

The charged dust particles confined in a plasma environment under gravitational and electrostatic forces interact with each other, usually via screened Coulomb (Debye-H\"{u}ckel) potential. Their behaviors are controlled by the Coulomb coupling parameter ($\Gamma$) and the screening constant ($\kappa$). A small number of such interacting dust particles under harmonic confinement provided by surrounding plasma may manifest in the formation of a dust cluster. The formation of dust clusters varying from one dimension to three dimensions in a plasma environment and their structures, and their properties in capacitively coupled RF discharge, are discussed by Melzer et al. \cite{melzer2010finite}. By suitably controlling the strength of vertical and horizontal confinements, they were successful in producing a 1D dust cluster. A zig-zag transition was also observed when the pressure was controlled externally. Sheridan and Wells \cite{sheridan2010dimensional} determined the critical exponents of such a zig-zag transition. Interestingly, Melzer et al. also produced $2$D finite dust clusters where the particles organize themselves into circular shells \cite{melzer2019finite}. They also observed $3$D spherical dust clusters, the so-called Yuakawa ball by suitably controlling the confining forces with a fewer ($N=22$) particles. Depending on the dominant interaction both spherical (in the presence of isotropic interaction) and chainlike structures (in the presence of attractive wakefield) may be possible. Note that the structure of dust clusters embedded in a plasma environment may be significantly influenced by the screening parameter $\kappa$ (and this introduces a difference of such a Yukawa cluster from the Coulomb cluster). Baumgartner et al. \cite{baumgartner2008ground} have studied the shell configuration of spherical Yukawa clusters in their ground states for different values of particle number and screening constant. Different properties of the Yukawa cluster such as cluster compression, change of average density profile, a transition from inner to outer shells, etc. were found to be influenced by screening for a given value of particle number. Thus, screening provides a different dimension to the Yukawa dust cluster compared to the Coulomb cluster. The micron-sized dust particles confined in plasma exhibit phase transition similar to solid-to-liquid transition in bulk matter. Nonequilibrium melting of $2$D finite dust cluster caused by instability was investigated by Ivanov et al. \cite{ivanov2005melting}. They observed a two-step transition from solid to hot crystalline state \cite{ichiki2004melting} with a reduction in discharge pressure resulting in unstable oscillations which then transit to fluid state on further reduction of gas pressure. Their results are in good agreement with the nonlinear simulations performed by Schweigert et al. \cite{schweigert1998plasma}. Melting transition in  finite $3$D clusters, so-called Yukawa balls, was experimentally studied by Schella et al.\cite{schella2011melting}. The angular correlation was found to decay before the vanishing of radial correlation. The critical value of the Coulomb coupling parameter was determined for a cluster containing $35$ dust particles.

The rotation of dust clouds is observed in several experiments of dusty plasma subjected to an external magnetic field. The rotation of a particle cloud  observed  by Sato et al. is attributed to the ion drag force on the fine particles \cite{sato2001dynamics}\cite{kaw2002rotation}. Konopka et al. reported that an external magnetic field results in a rotation of dust clouds suspended in the sheath of a radio-frequency discharge \cite{konopka2000rigid}. They suggested an analytical model that explains qualitatively the mechanism of particle rotation, which depends on electrostatic force, ion drag, neutral drag, and effective interparticle interaction forces. Interestingly, intershell rotation of dust particles in two dimensions was also reported by Maity et al. \cite{maity2020dynamical} in the absence of a magnetic field which they attributed to the unbalanced electric force between the inner and outer shells.

The study of the behavior of dusty plasma in presence of an external magnetic field may be of profound interest from the point of view of laboratory, fusion plasma, for various industrial applications as well as interstellar, and solar plasma environments. Single dust particle rotation in dc glow discharge plasma in the presence of a magnetic field was observed by Karasev et al. which they attributed to the impulse exerted in tangential direction by the plasma flux on a particle \cite{karasev2009single}. Recent experiments on dusty plasma under the influence of an external magnetic field have revealed various interesting properties of such a system. Ordered structures imposed on dusty plasma systems have been observed at high magnetic field strength in a magnetized dusty plasma experiment (MDPX) \cite{thomas2015observations}. Dust waves and plasma filamentation have also been observed in the MDPX facility \cite{thomas2016initial}. While a  low- to moderate-strength magnetic field may influence the dust charging and structure formation via the plasma particle dynamics, a large magnetic field of sufficient strength may have a direct influence on grains which leads to modification in the transport properties of dust through plasma as well as the formation of structures.  Dust is an important constituent of the interstellar medium. The coupling of dust with the magnetic field may play a very important role in stellar dynamics \cite{beitia2017interstellar}. It is known that dust grains grow by accretion and coagulation in dense environments. The study of dust clusters may be of immense importance in such environments. Hirashita \cite{hirashita2012dust} has investigated the impact of dust growth on the extinction curve. The purpose of the present study is to investigate the effect of magnetic field and temperature on the dynamical behavior of finite-sized charged dust particles in confined geometry. 

\section{The Model}

A three-dimensional dusty plasma containing $N$ dust particles in the background of quasi-neutral plasma confined in a box is considered. The dust grains are assumed to interact among themselves via the repulsive Debye-H\"{u}ckel potential,
\begin{equation}
	V(r)=\frac{q_{d}}{4\pi\epsilon_{0}r}\exp(-r/\lambda_{d}),
	\label{Debye-Huckel_Poten}
\end{equation}
where $q_{d}$ is dust charge, $\lambda_{d}$ is dust Debye length and $r$ is the inter-particle distance between two dust grains. The dust Debye length is obtained from the ion and electron Debye lengths as $\lambda_{d}=\frac{\lambda_{e}\lambda_{i}}{\sqrt{\lambda_{e}^{2}+\lambda_{i}^{2}}}$, where $\lambda_{e}$ and $\lambda_{i}$ are electron and ion Debye lengths respectively and are defined as $\lambda_{e}=\sqrt{\frac{\epsilon_{0}k_{B}T_{e}}{n_{e}e^{2}}}$ and $\lambda_{i}=\sqrt{\frac{\epsilon_{0}k_{B}T_{i}}{n_{i}e^{2}}}$. Here, $k_B$ is the Boltzmann constant, and $T_e, n_{e}$ and $T_i, n_{i}$ are the temperature and number density of electron and ion respectively. As a model for confinement,  the isotropic harmonic potential is considered \cite{arp2005confinement, ludwig2005structure}. The confining harmonic potential is assumed to represent the superposition of gravitational, thermophoretic, electric field, and ion drag force acting on the dust particles. In addition, the system is subjected to an external magnetic field along the $z$ direction. Thus, the system can be considered as some charged particles in an electromagnetic field. Then, the Hamiltonian of this system is \cite{goldstein2002classical}
\begin{equation}
	H=\frac{1}{2m}\sum_{i=1}^{N}(\mathbf{p_{i}}-q_{d}\mathbf{A}_{i})^2+q_{d}\phi,
	\label{Hamiltonian_Eq}
\end{equation}
where 
\begin{equation}
	q_{d}\phi=\frac{q_d^2}{4\pi\epsilon_0}\sum_{i=1}^{N-1}\sum_{j=i+1}^{N}\frac{\exp(-r_{ij}/\lambda_{d})}{r_{ij}}+\frac{1}{2}m \omega^{2} \sum_{i=1}^{N} r_{i}^2.
\end{equation}
Here, $r_{i}$ is the distance of the $i$th particle from the center of the box and $r_{ij}=|\mathbf{r_i}-\mathbf{r_j}|$. $m$ and $q_{d}$ are mass and charge of a dust particle respectively and $\lambda_{d}$ is the Debye length of the dust grains. $\omega$ denotes the strength of the confinement potential. The equation of motion of the $i$th particle is
\begin{equation}
	m\ddot{\mathbf{r}}_i=q_d(\mathbf{v}_i\times\mathbf{B})-q_{d}\mathbf{\nabla}\sum_{j\neq i}^{N} V(r_{ij})-m\omega^{2}\mathbf{r}_{i},
	\label{EOM_F}
\end{equation}
where $q_{d}(\mathbf{v_{i}}\times\mathbf{B})$ is the Lorentz force experienced by the charged dust particle due to the magnetic field and $V(r_{ij})$ is the Debye-H\"{u}ckel potential operative among the dust grains, and the last term represents the confining harmonic force.   

We further recast Eq. (\ref{EOM_F}) into dimensionless form using the  scaled variables $\mathbf{r}^\prime = \mathbf{r}/\lambda_d$, $\tau = \sqrt{\frac{k_BT_d}{m\lambda_d^2}}t$, $\mathbf{B}^\prime = \frac{q_d \lambda_d}{\sqrt{mk_BT_d}}\mathbf{B}$, and $\Omega^2= \frac{m\lambda_d^2}{k_BT_d}\omega^2$. Then, the dimensionless equation of motion in terms of scaled variables reads
\begin{equation}
	\ddot{\mathbf{r}_i^\prime}=(\mathbf{v}_i^{\prime}\times\mathbf{B^{\prime}})+\Gamma\kappa\sum_{j\neq i}^{N}\frac{\Big[1+{r_{ij}^{\prime}}\Big]}{{r_{ij}^{\prime}}^{3}}\exp\Big(-{r_{ij}^{\prime}}\Big){\mathbf{r_{ij}^{\prime}}}-\Omega^{2}{\mathbf{r_{i}^\prime}}.
	\label{EOM_Scaled}
\end{equation} 
The overdot now refers to the redefined time derivative. $T_d$ denotes the dust kinetic temperature and $\Gamma$ and $\kappa$ are known as the Coulomb coupling and screening parameters respectively, which are defined as  
\begin{equation}
	\Gamma=\frac{q_{d}^{2}}{4\pi\epsilon_{0}r_{av}k_{B}T_{d}}
	\label{Coupling-Param}
\end{equation}
and
\begin{equation}
	\kappa=\frac{r_{av}}{\lambda_d}.
	\label{Screening-Param}
\end{equation}
Here, $r_{av}$ is the average interparticle distance of the charged dust particles, defined as $r_{av}=(\frac{3}{4\pi n_{d}})^{1/3}$ and $n_{d}$ is the number density of the dust particles.  $\Gamma > 1$ indicates that the average interparticle interaction energy dominates over average thermal energy and the system is said to be in a strongly coupled state, whereas $\Gamma < 1$ refers to a weakly coupled state\cite{hamaguchi1994thermodynamics}. Note that the effect of temperature on the dynamics of the dust particle is studied through the Coulomb coupling parameter.

\section{Simulation scheme}

 Molecular dynamics simulation was performed on  $32$ charged dust particles placed inside a cubical simulation box, interacting via Debye-H\"{u}ckel potential [see Eq. (\ref{Debye-Huckel_Poten})]. The size of the simulation box is chosen as $L_{x}=L_{y}=L_{z}=6.83\times10^{-4}$m. A modified version of the velocity-verlet algorithm was used to integrate the equations of motion \cite{spreiter1999classical}. In the simulation, the values of ion, electron, and dust number densities respectively are  $n_i=10^{15} \;m^{-3}$, $n_e=8.89\times10^{14}\;m^{-3}$, and $n_d=10^{11}\;m^{-3}$ and the electron and ion temperatures respectively are $T_e=2320\; K$ and $T_i=2050 \; K$. The mass of the dust particles is taken to be $m=6.99\times10^{-13}\;kg$. The value of the screening parameter is calculated and is kept fixed at $\kappa=1.8$ for all the runs. The frequency of the harmonic potential is fixed at $\omega=50~Hz$.

A simulation run starts from a random initial configuration of the particles. For each run,  the number of particles, volume, and temperature are kept fixed. To simulate at a fixed temperature, a Berendsen thermostat \cite{berendsen1984molecular,  morishita2000fluctuation} is used. For the initial $1.4\times10^{6}$ steps the system is coupled to the Berendsen thermostat to bring the system to equilibrium at the desired temperature and data is collected for the next $1.0\times10^{5}$ steps.

\section{Results and Discussion}
\label{RandD}
The central objective of the present investigation is to study the dynamics and phase transition of the Yukawa dust cluster under the influence of an external magnetic field.  The simulation is performed with $N=32$ point-sized charged dust particles.  The physically relevant parameters in our model are coupling parameter $\Gamma$,  the applied magnetic field $B$, screening constant $\kappa=r_{av}/\lambda_d$, and the frequency of the harmonic potential $\omega$. To reduce the
number of free parameters, we fixed the screening constant $\kappa=1.8$ and frequency $\omega=50~Hz$. While temperature or the Coulomb coupling constant $\Gamma$ controls the kinetic energy of the charged dust particles, the effective dynamics are controlled by the applied magnetic field $B$. To understand the dynamics of the system of dust particles with the coupling strength and the applied magnetic field, the representative parameters $\Gamma$ and $B$ are varied for a wide range of values. For the current investigation, the magnetic field $B$ is varied in the range $0.001-0.7~T$. On the other hand, coupling parameter $\Gamma$ is varied by changing the dust temperature in the range $293 - 40000\; K$. Anomalously high dust kinetic temperature is reported in several dusty plasma experiments \cite{thomas2010driven}\cite{fisher2013thermal}\cite{williams2007measurement}. The interplay between repulsive Debye-H\"{u}ckel potential and the confining harmonic potential results in a shell-like arrangement of the cluster of dust particles. 
\begin{figure}[!b]
	\centering{
		\includegraphics[width=6.5cm,height=4.2cm]{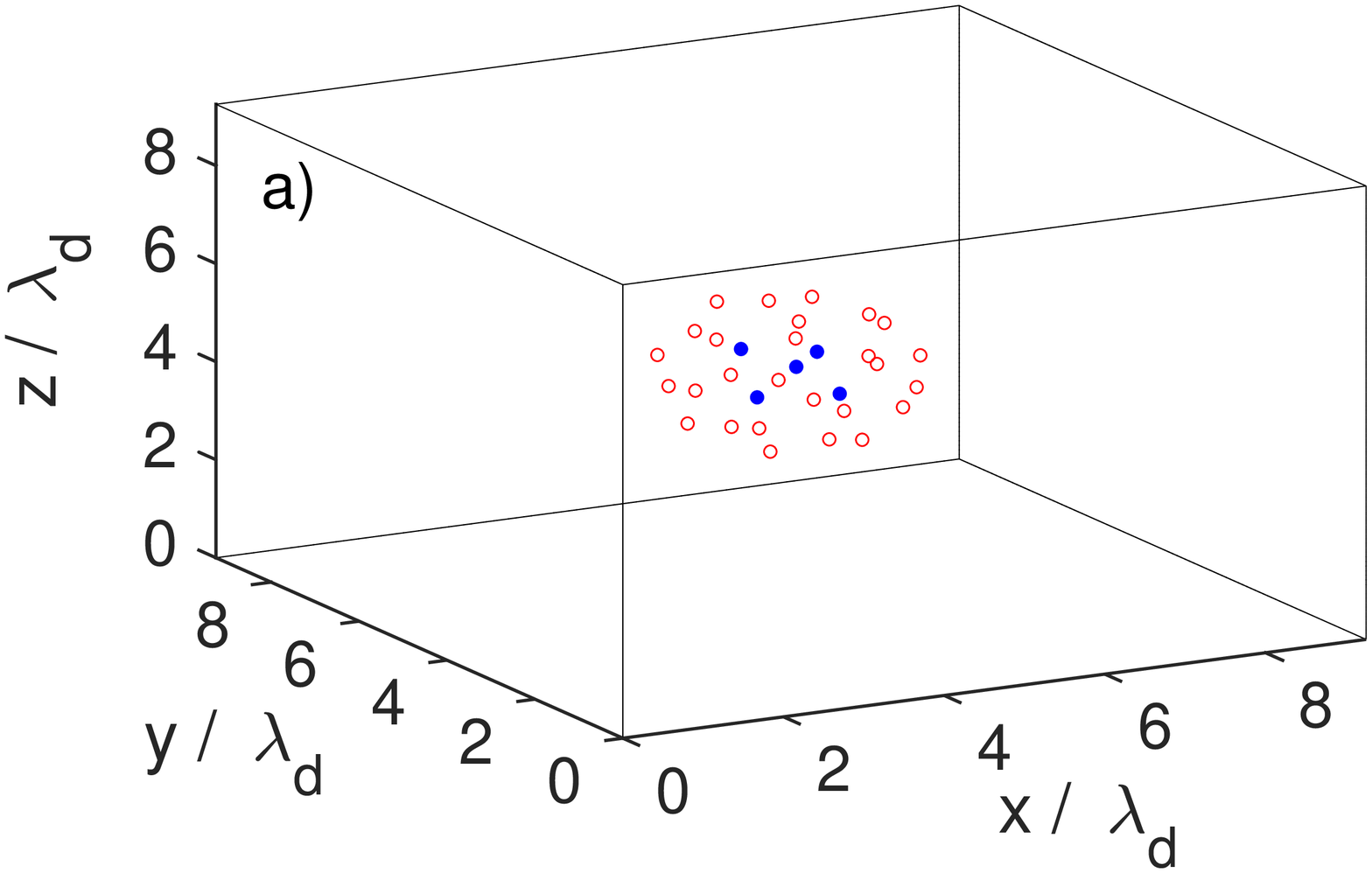} 
		\includegraphics[width=6.5cm,height=4.2cm]{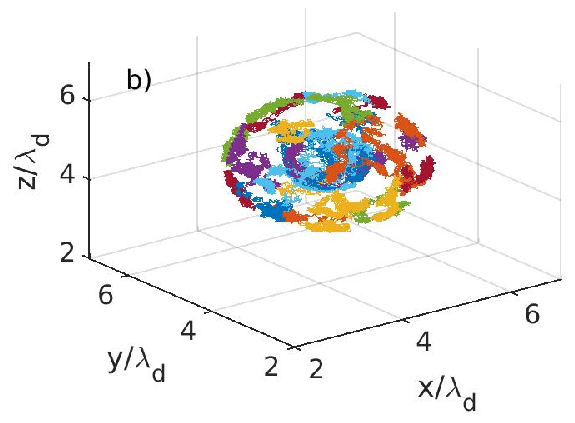}
		\includegraphics[width=6.5cm,height=4.2cm]{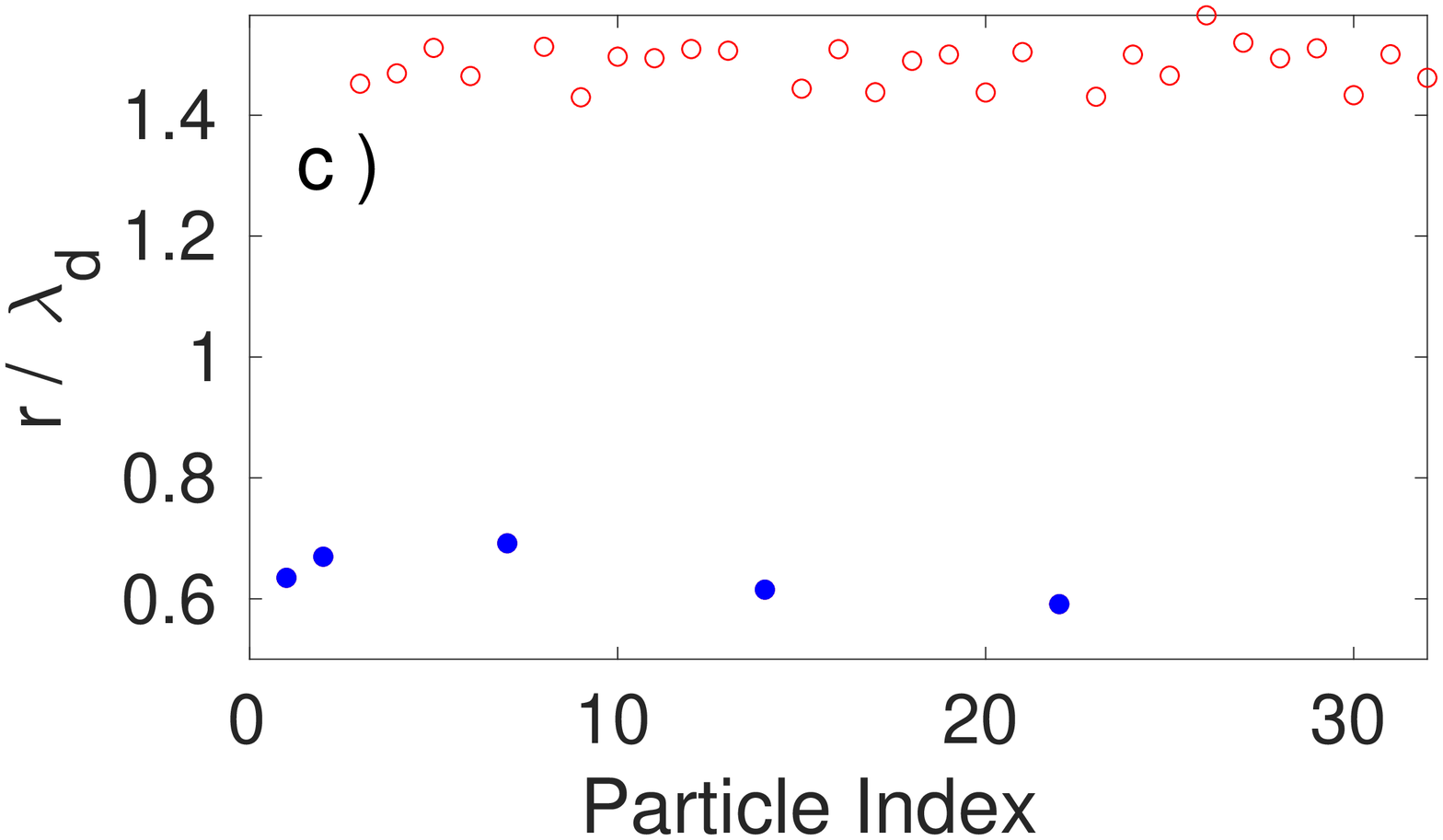}
		}
	\caption{(Color online) Organization of the charged dust particle cluster containing $N=32$ particles at $\Gamma=514.38$ and $B=0\;T$. (a) A snapshot of the shell configuration of the dust cluster. The filled circles represent particles on the inner shell, and the open circles represent the outer shell particles. (b) The trajectories of the particles for $1.0\times10^{5}$ time steps. (c) The radial position of the particles  measured from the center of the simulation box at the final time step of the simulation. }
 	
	\label{Trajec-1}
\end{figure}
For the parameters used here, charged dust particles organize themselves into two nested shells with a configuration ($5$, $27$). This configuration remains invariant under any change in the magnetic field. Figure \ref{Trajec-1}(a) shows a snapshot of the dust particle cluster and Fig. \ref{Trajec-1}(b) depicts the time evolution of the particle trajectories for $\Gamma=514.38$ at zero magnetic fields. The organization of the particles into two shells can be seen in Fig. \ref{Trajec-1}(c) which shows the radial position of the particles measured from the center of the simulation box.

\subsection{Dynamics of the charged particles as a function of magnetic field and Coulomb coupling parameter}
\begin{figure}[!bht]
	\vbox{
		\includegraphics[width=8cm,height=4.5cm]{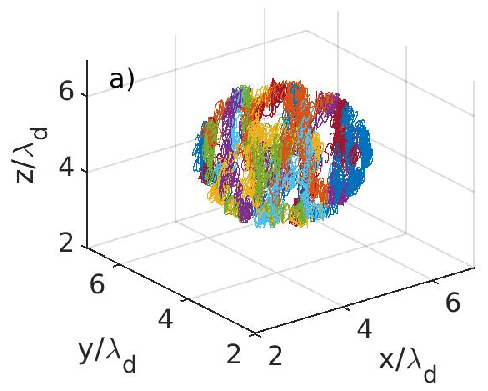}
		\includegraphics[width=8cm,height=4.5cm]{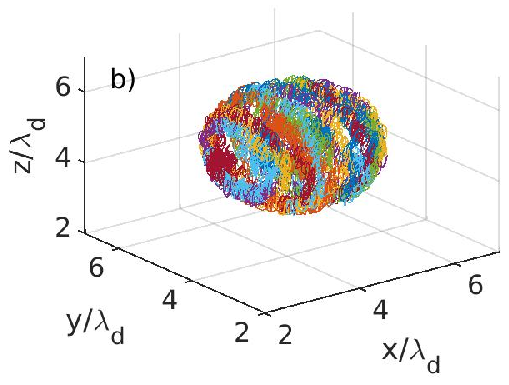}
		\includegraphics[width=8cm,height=4.5cm]{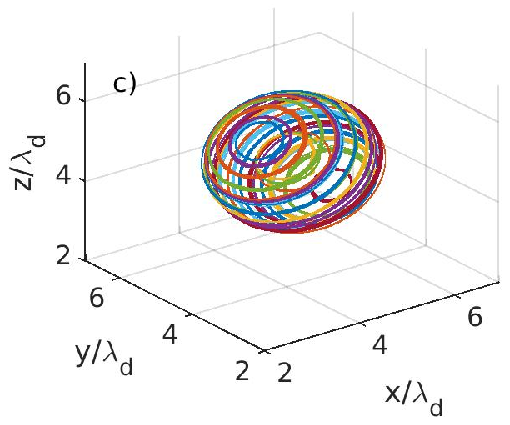}
	}
	\caption{(Color online)  Trajectories of the dust particles for three different values of the applied magnetic field strengths (a) $B=0.001~T$, (b) $B=0.4$ T, and (c) $B=0.6$ T, for $\Gamma=21.11$.}
	\label{B-Vari}
\end{figure}

To understand the equilibrium properties of the charged dust particles as a function of the magnetic field, the Coulomb coupling constant is fixed and the dynamics of the particles as a function of the applied magnetic field $B$ were studied. The particle trajectories for different values of magnetic field $B$ are shown in Fig. \ref{B-Vari}. The charged dust particles exhibit interesting dynamics as a function of the magnetic field, with the particles being organized into two distinct shells as shown in Fig. \ref{Trajec-1}(a). 
Since the dust particles are charged particles and the magnetic field is applied along the $z$ direction, the dust particles will experience a Lorentz force and start rotating about the $z$ axis. However, in contrast, at a low applied magnetic field $B=0.001~T$, the mean trajectories of the charged dust particles show rotation about a random orientation. Considerable fluctuation about the mean trajectory is evident in Fig. \ref{B-Vari}(a). The trajectory plot suggests that the charged dust particles attain vibrational motion along with rotational motion. The rotational motion is due to the external magnetic field, and the vibrational motion about the mean can be attributed to the thermal energy. As the magnetic field increases, at $B=0.4~T$, the system of particles tries to attain a definite axis of rotation. However, the fluctuation in the mean trajectory of a particle attributed to vibrational mode due to thermal energy remains. With further increase of the magnetic field to $B=0.6~T$,  a drastic change in the dynamics of the system of particles is observed. At this magnetic field strength, the vibrational motion of the trajectories around the mean trajectories of the system of particles completely ceases, leaving only the rotational motion. Thus, the charged dust particles show a phase transition from disordered rotation to ordered rotational motion. The system of dust particles has collectively developed a phase where all the particles rotate about a distinct axis.

The fact that the collective dynamics of the dust particles at a given coupling parameter changes drastically as a function of the applied magnetic field induced us to explore the possibility of the effect of coupling strength. 
\begin{figure}[h]
	\vbox{
		\includegraphics[width=8cm,height=4.5cm]{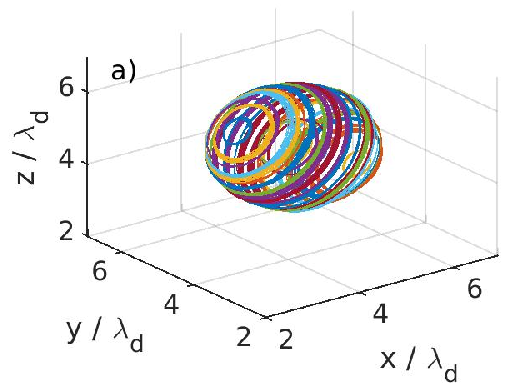}
		\includegraphics[width=8cm,height=4.5cm]{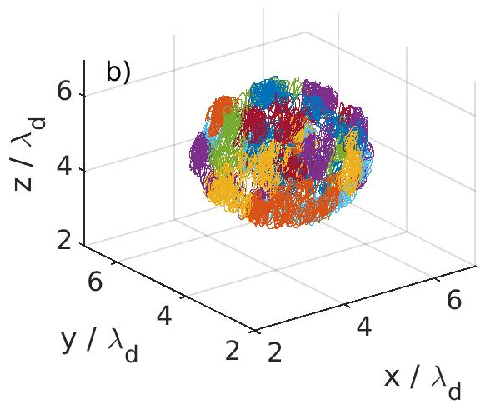}
	}
	\caption{(Color online)  Trajectories of the dust particles at two different coupling strengths (a) $\Gamma = 15.43$, (b) $\Gamma = 12.86$ keeping magnetic field fixed at $B=0.6$ T.}
	\label{T-Vari}
\end{figure}
Thus,  the coupling parameter was varied, keeping the magnetic field fixed. The representative trajectory for two different values of coupling parameter keeping the magnetic field fixed at $B=0.6~T$ is shown in Fig. \ref{T-Vari}(a) and \ref{T-Vari}(b).  It is observed that, at low coupling strength for a fixed magnetic field, the dust particles have both rotational as well as vibrational modes. The dust particles acquire vibrational motion about the mean trajectory along with the rotational motion. However, the vibrational symmetry is broken at high values of coupling strength, and the charged dust particles homogeneously rotate about a fixed axis; i.e., the system has made a transition from high symmetry to a low symmetry phase, indicating a phase transition. The change in dynamics of the dust particle is rationalized by competing length and time scales in the system. At a high value of Coulomb coupling parameter and high magnetic fields, in equilibrium, the force due to repulsive Debye-H\"{u}ckel potential balances with the attractive harmonic force by minimizing the average interparticle distance between the charged dust particles. The total kinetic energy is then converted to rotational energy by the magnetic field. Thus, the system of particles attains a fixed axis of rotation. At smaller values of the Coulomb coupling parameter, the equilibrium repulsive potential intersects the harmonic potential at two points, and the trajectories of the particles are confined between these two equilibrium energy shells. The low magnetic field is insufficient to convert the entire kinetic energy into rotational energy. Thus, the system of particles keeps switching itself between these two equilibrium energy shells that give rise to the vibrational mode of the particles. The above analysis of the trajectories of the particles suggests the possibility of a phase transition as a function of the control parameters i.e., coupling parameter $\Gamma$ and magnetic field $B$. 

\subsection{Phase transition}

Phase transitions represent singularities in the free energy functional as the control parameter of the system is varied. In a macroscopic system, the Lindemann parameter, defined as particle position fluctuation normalized by interparticle distance, or relative interparticle distance fluctuation (IDF), shows a sudden jump during melting. The IDF is defined mathematically as \cite{schella2011melting}
\begin{equation}
	u_{rel}=\frac{2}{N(N-1)}\sum_{i=1}^{N-1}\sum_{j=i+1}^{N} \sqrt{\frac{<r_{ij}^{2}> - <r_{ij}>^{2}}{<r_{ij}>^{2}}}.
	\label{Rel-Fluc}
\end{equation}
However, the discontinuous change in the IDF that characterizes a phase transition is hard to observe in a finite system. Since the number of particles in our system is limited to ($N=32$) small numbers, we adopt the strategy suggested by Boning et al. \cite{boning2008melting} to identify order-disorder transitions in finite-size systems. The variance of block-averaged interparticle distance fluctuation (VIDF) is considered to be a promising diagnostic tool for identifying transition points. The VIDF serves as an alternate representation of the order parameter of macroscopic systems in finite-size systems.  It is calculated by first dividing the simulation duration in equilibrium into a certain number of blocks ($M$) of equal duration and calculating the IDF $u_{rel}$ for each block. Then the VIDF is defined as,
\begin{equation}
	\sigma =<u_{rel}^{2}>-<u_{rel}>^{2},
	\label{VIDF_Eq}
\end{equation}
where
\begin{eqnarray*}
    \nonumber
	<u_{rel}^{2}>&=&\frac{1}{M}\sum_{\alpha=1}^{M} u_{rel}^{2}(\alpha), \\
    <u_{rel}> & = & \frac{1}{M}\sum_{\alpha=1}^{M} u_{rel}(\alpha).
\end{eqnarray*}
Boning et al. demonstrated that the VIDF exhibits a distinct peak during the melting transition in a finite-size system. Thus, the identification of transition points is very efficient. To identify the critical values of the magnetic field and the coupling constant, one parameter is kept fixed and the other one is varied. At first, the Coulomb coupling parameter $\Gamma$ is fixed, and the VIDF is calculated over a wide range of magnetic fields. 
\begin{figure}[h]
		\includegraphics[width=8cm,height=4.0cm]{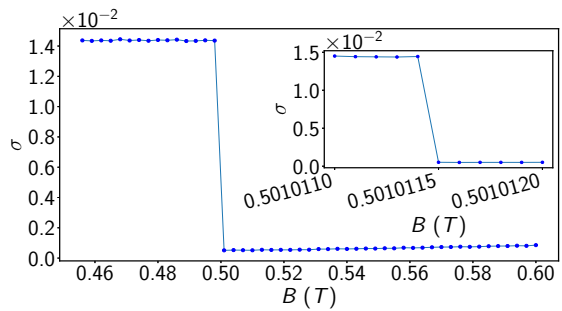}
	\caption{(Color online) Variation of the variance of block averaged inter-particle distance fluctuation ($\sigma$) to the strength of the magnetic field $B$ for fixed $\Gamma=21.107216$.}
	\label{VIDF-vs-B}
\end{figure}
Fig. \ref{VIDF-vs-B} shows the variation of the VIDF with change in the magnetic field strength for $\Gamma=21.107216$. The inset shows closer scanning of the magnetic field. It is seen that at around $B\sim0.5~T$, there is an abrupt discontinuity in the VIDF, indicating a singularity. Investigating the trajectories of the charged dust particles, it is found that the system of particles undergoes a phase transition from a disordered rotating phase to an ordered rotating phase, by breaking the vibrational symmetry of low magnetic field strength. The value of the critical magnetic field for this transition is $B_c=0.5010114~T$ at  $\Gamma=21.107216$ as VIDF abruptly drops to a lower value on a slight increase of field strength beyond this value.

\begin{figure}[!bht]
		\includegraphics[width=8cm,height=5cm]{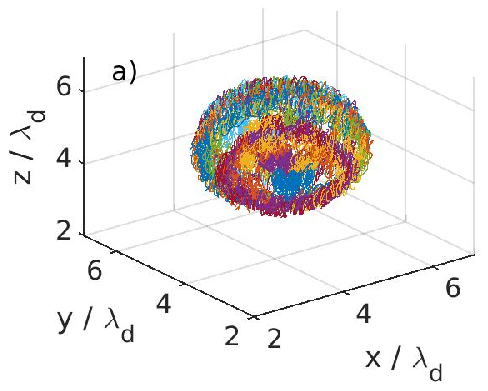}
		\includegraphics[width=8cm,height=5cm]{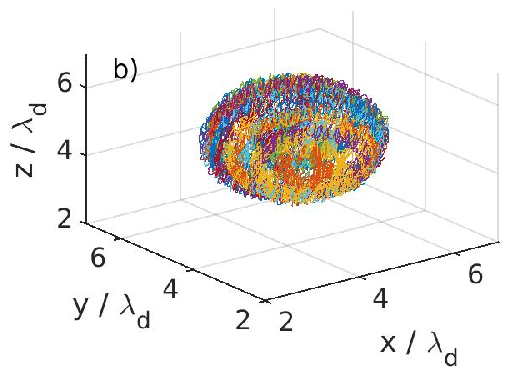}
		\includegraphics[width=8cm,height=5cm]{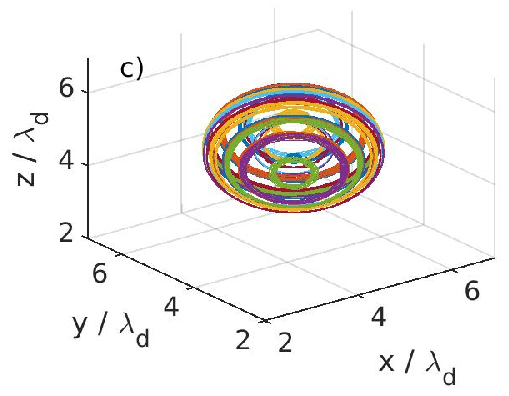}
		\caption{ (Color online) Plot of trajectories of the dust particles at the critical magnetic field strength $B_c=0.5010114~T$, for three different coupling strengths (a) $\Gamma = 21.107211$, (b) $\Gamma_c=21.107216$, and (c) $\Gamma=21.107219$}
	\label{Trajectories-Tvari}
\end{figure}
\begin{figure}[h]
\centering{
		\includegraphics[width=8cm,height=4.0cm]{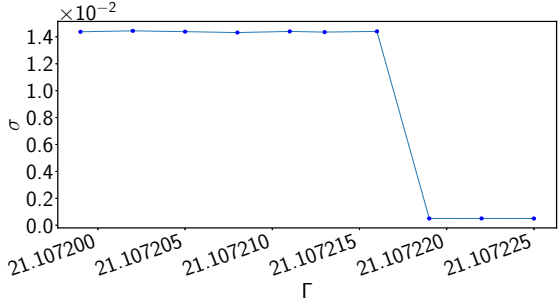}}
	\caption{(Color online) Variation of the VIDF with coupling parameter at the critical magnetic field $0.5010114~T$.}
	\label{VIDF-vs-T}
\end{figure}
\begin{figure}[h]
\vbox{
\centering{
\includegraphics[width=7.5cm,height=4.8cm]{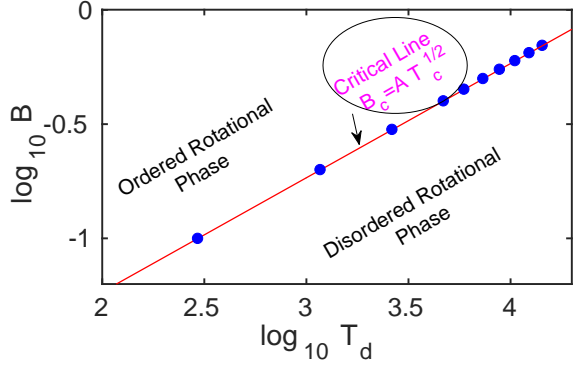}

		}
		}
	\caption{(Color online) The $B-T$ phase diagram separating ordered-to-disordered rotational phase of the finite-sized system of dust particles. The circles indicate numerically obtained data points and the straight line is the critical line satisfying the equation of state $B_c= A\sqrt{T_{c}}$ with $A = 5.8\times 10^{-3}\; TK^{-1/2}$.}
	\label{BT-PhaseDiagram}
\end{figure}

To see if the transition of the dynamics of the system of the finite number of dust particles is true, the effect of coupling strength around this critical magnetic field is also investigated. By keeping the magnitude of the magnetic field fixed at $B_c=0.5010114~T$,  the coupling strength of the dust particles is varied. The dynamics of the system of particles changes abruptly as the coupling strength slightly deviates from the critical value $\Gamma_{c} = 21.107216$ (Fig. \ref{Trajectories-Tvari}). Further increase in the coupling strength of the dust particles does not induce any change in the dynamics. This study demonstrates that the system of dust particles shows distinct collective dynamics at the critical point ($B_c$, $\Gamma_{c}$). The response of the system beyond the critical point is drastically different, suggesting the emergence of two different phases around the critical points.  This exercise suggests that the system undergoes a first-order phase transition at the critical point ($B_c$, $\Gamma_c$). To see if the transition is captured by the VIDF ($\sigma$), the VIDF as a function of the coupling strength $\Gamma$ is plotted at the critical magnetic field $B_c=0.5010114~T$, as shown in Fig. \ref{VIDF-vs-T}.  
The discrete drop at the critical value of the coupling strength, $\Gamma_c=21.107216$, clearly indicates that the system undergoes a first-order phase transition at this critical point. Further increase in the coupling parameter does not induce any change in the VIDF, suggesting that the system behaves collectively above the critical point. Using the VIDF as the signature of this first-order phase transition from the ordered to the disordered rotational phase, the phase diagram in the $B-T$ plane (since $\Gamma \propto 1/T_d$) separating the two phases is shown in Fig. \ref{BT-PhaseDiagram}. 
The critical magnetic field ($B_c$) corresponding to a dust temperature ($T_c$) or coupling strength ($\Gamma_c$) shows a power-law dependency $B_c=AT_c^\alpha$ at the fixed dust density $n_d$ with exponent $\alpha=0.5$ and $A = 5.8\times 10^{-3}\;TK^{-1/2}$. Thus, the phase boundary separating the two phases at a fixed dust density and finite dust temperature follows the equation of state
\begin{equation}
\frac{B_c}{\sqrt{T_c}} = \text{const}.
	\label{E-o-F} 
\end{equation}
The emergence of the square root dependence of the critical magnetic field on the dust temperature is easy to understand from the following physical picture. The order-to-disorder transition is characterized by the onset of coherent rotation about an axis when the magnetic field is applied. The overlapping of the trajectories of the particles in the disordered state disappears and the particles rotate in distinct well-defined trajectories. The kinetic energy due to the available thermal energy of the dust particles is then completely converted into rotational energy by the magnetic field. For coherent rotation in thermal equilibrium, each charged dust particle rotates about the fixed axis of rotation with an angular velocity  $\dot \theta$. Then,  the available thermal kinetic energy $\sim k_BT_c$ equals the rotational energy $I \dot \theta^2/2$ of the dust particle, which may be the superposition of rotational energy of cyclotron motion with angular velocity $\frac{q_dB_c}{m}$  and a rotational drift suggesting $B_c \propto \sqrt{T_c}$. Below the critical magnetic field $B_c$, for a fixed dust temperature $T_d$, the available thermal energy decomposes into rotational and linear kinetic energy. The excess linear kinetic energy then gives rise to the vibrational motion around the mean, giving rise to the disordered rotational phase.

\subsection{Structure dynamics of dust particles}
To study the structural properties of the dust clusters a radial distribution function was used. The radial distribution function (RDF) is proportional to the probability of finding a pair of particles separated by a distance in the range $r$ to $r+dr$ from a reference particle, and is defined as \cite{hansen2013theory} 
\begin{equation}
    g(r)=\frac{1}{N}\Big<\sum_{i}^{N} \sum_{j\neq i}^{N} \delta(r-r_{ij})\Big>.
    \label{RDF-Eq}
\end{equation}
It gives an idea of how the particles arrange themselves around one another.

To get an insight into the structure of the dust cluster with coupling strength $\Gamma$,  RDF  was calculated for a range of the Coulomb coupling parameter $\Gamma$, initially for a  magnetic field $B=0\; T$. An analysis similar to the previous section suggests that at $\Gamma_c=314.67$ a transition from a disordered state to an ordered state takes place. This can also be seen from the plot of $g(r)$.
\begin{figure}[!htb]
	\centering{
		\includegraphics[width=8cm,height=5.0cm]{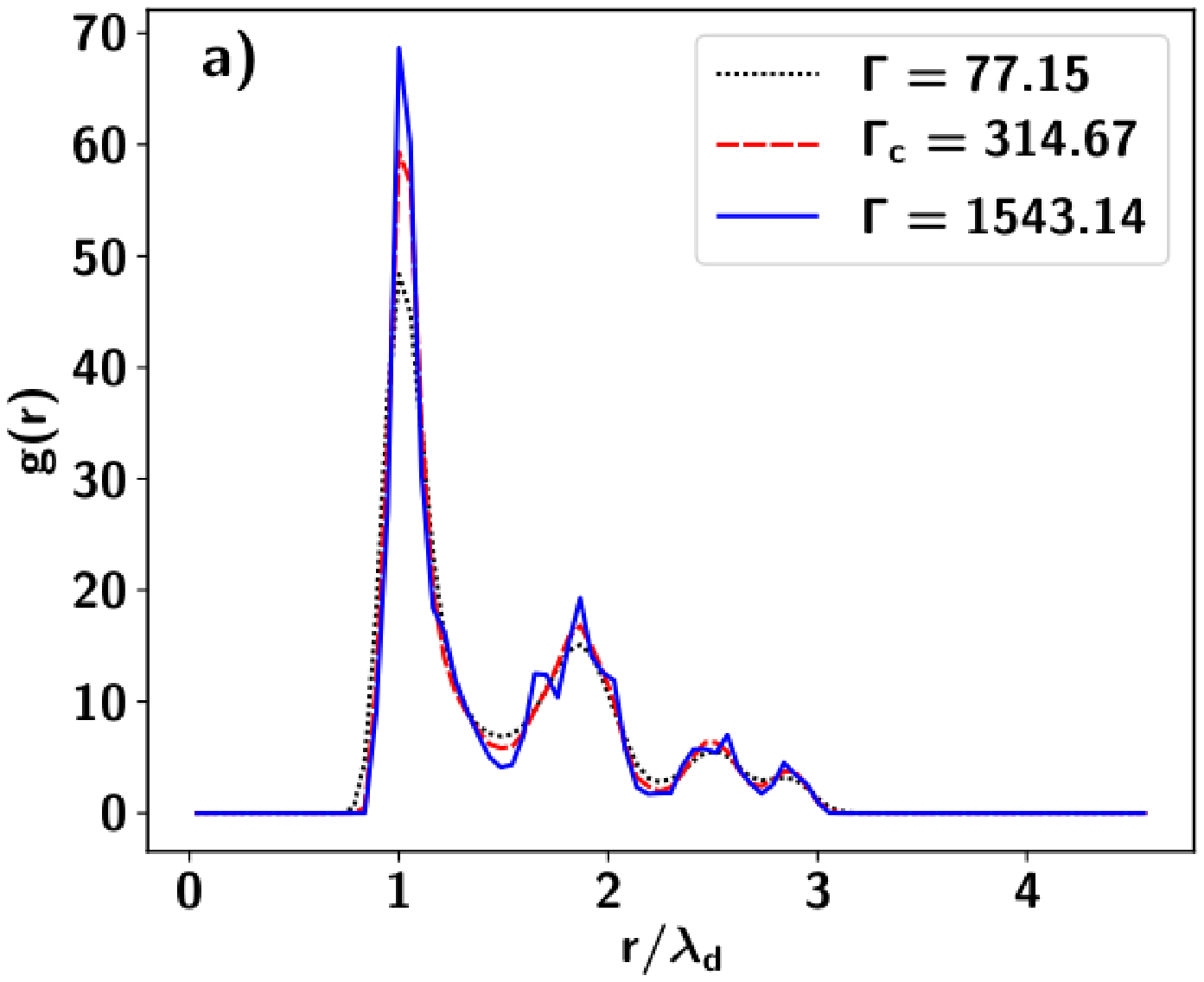}
		\includegraphics[width=8cm,height=5.0cm]{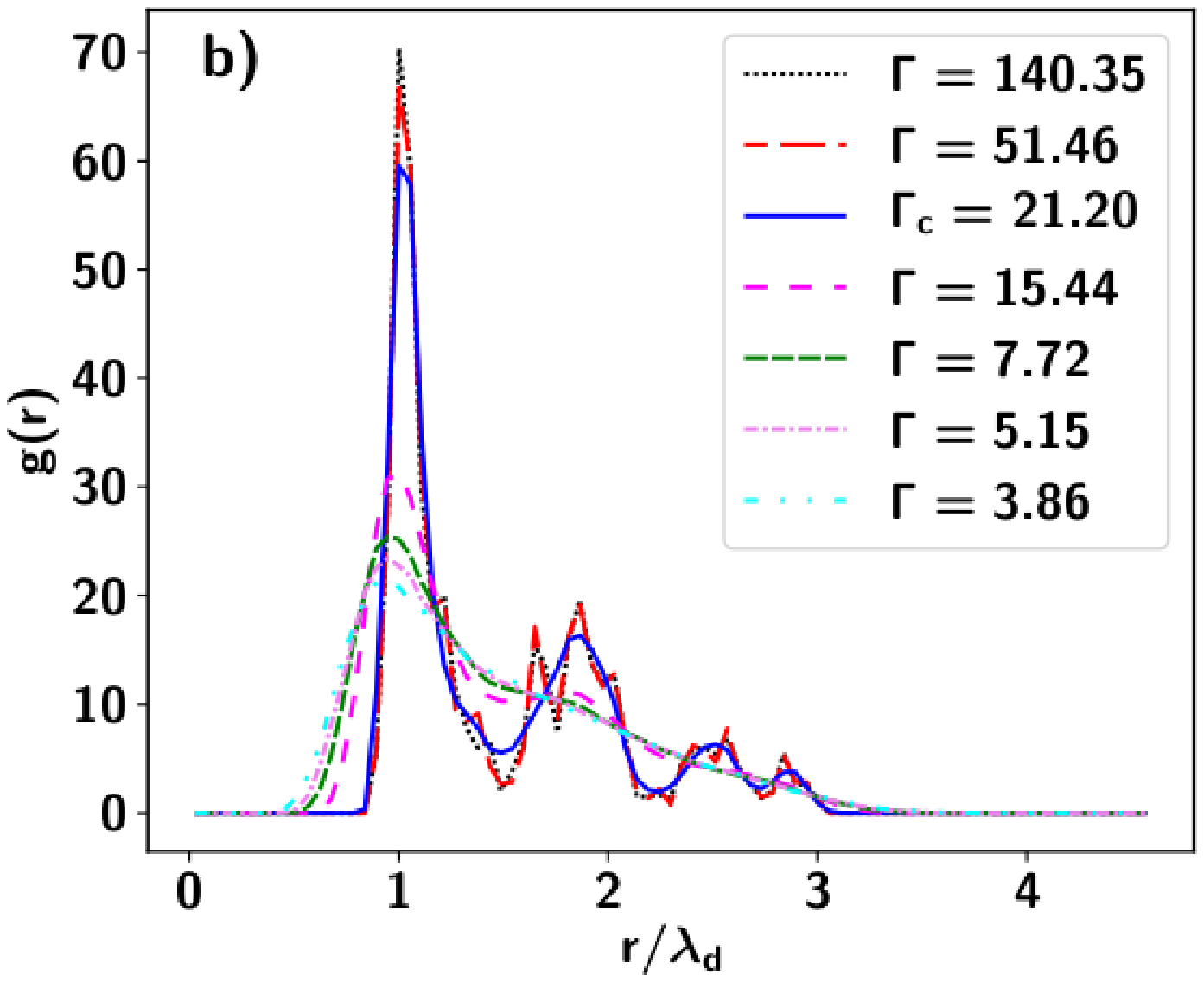}
		}
	\caption{(Color online) The radial distribution function $g(r)$ with $\Gamma$ at (a) $B=0\;T$, (b) $B=0.5~T$.}
	\label{RDF-T}
\end{figure} 
The plot of the RDF for different values of $\Gamma$ at a  magnetic field $B=0~T$  is shown in Fig. \ref{RDF-T}(a). For  $\Gamma=77.15$, the RDF suggests a liquidlike structure. However, the height of the first peak of $g(r)$ is seen to increase with increasing value of coupling parameter $\Gamma$ from $77.15$ to $1543.14$, and there is partial development of secondary peaks beyond $\Gamma_c=314.67$ suggesting correlation, as in a partially crystallized state. On the other hand, at $B=0.5$ T of Fig. \ref{RDF-T}(b), the height of the first peak initially increases gradually from $3.86$ to $15.44$ and then exhibits a sudden change corresponding to $\Gamma=21.20$. The sharp peaks of $g(r)$ beyond the $\Gamma=21.20$ suggest a strong correlation among the particles at large distance reminiscent of the fixed position of particles, as in a solid. Furthermore, for the values of $\Gamma=51.46$ and $\Gamma=140.35$, the RDFs are almost superimposed which suggests the structure of the dust particles remains invariant with the change.  The sudden change in the peak height and development of subsequent peaks of $g(r)$ is a signature of the transition from a disordered to an ordered state.

\section{Summary and Conclusions}
In summary,  the dynamics of a finite-sized charged dust cluster under a confining harmonic and repulsive Debye-H\"{u}ckel potential subjected to an external magnetic field was studied. In the absence of the applied magnetic field at a finite value of coupling strength, the charged dust particles organize themselves into two nested spherical shells without any long-range order, similar to a fluidlike state. At very low coupling strength, the dust particles organize randomly, similarly to particles in their gaseous states. As soon as the magnetic field is turned on, at a weak magnetic field and small coupling parameter (or high temperature), the dust particles rotate around the surface of the two nested spheres in random order without any definite axis of rotation. The particles in this state exhibit both rotational and vibrational motions. The radial distribution function in this state reveals that the particles are organized randomly on the sphere. At a high magnetic field for a fixed coupling strength, a collective mode emerges and the dust particles rotate in order about a definite axis of rotation. This collective mode emerges by breaking the vibrational symmetry of a low magnetic field. Interestingly, the collective mode also can be achieved by tuning the coupling parameter or the temperature of the dust particles for a fixed magnetic field. This gives us a phase boundary between the orientationally ordered rotating fluid and the disordered rotating fluid phase that satisfies an equation of state $B_c/\sqrt{T_c}=$\;const at a constant dust density and finite temperature. The square root dependence of the critical magnetic field on the dust temperature is attributed to the available thermal energy converted to rotational energy by the applied magnetic field.  

Our analysis shows that the system of dust particles inherently organizes itself in spherical shells irrespective of the strength of the magnetic field, which can be attributed to the dynamical equilibrium between the attractive and repulsive potentials. This study may be useful in determining the magnetic field dynamics of stars in their early formation. However, the microscopic mechanism behind the rotation of dust clusters in the presence of a magnetic field is still an open question. In most of the experiments, the observed rotation is explained based on the ion drag model \cite{konopka2000rigid}\cite{sato2001dynamics}\cite{karasev2006rotational}. In contrast, Cheung et al. pointed out that the estimated value of ion drag force required for the observed rotation in their experiment of dust clusters in an inductively coupled rf plasma in the presence of an external magnetic field was much lower, and suggested that it cannot be fully responsible for the rotational motion\cite{cheung2003rotation}.
In the present work, we focus mainly on the dust dynamics in the presence of repulsive interparticle Yukawa interaction, confining potential, and Lorentz force due to an external magnetic field (ignoring the effect of dust-neutral collision and ion dynamics). It is interesting to see that, even in the absence of ion dynamics and related $\mathbf{E}\times\mathbf{B}$ drift, the dust cluster exhibits coherent rotation once the magnetic field exceeds a critical value. This rotational motion may have its origin in the coupling of Lorentz force and residue of Yukawa and harmonic forces. A more in-depth study will be required for a complete understanding of this process and is presently under study.

A preliminary investigation into the microscopic origin of the ordered-to-disordered phase transition points to chaotic dynamics of particles. This work is currently in progress.

\section*{acknowledgement}
H.S. gratefully acknowledges financial support by Tezpur University under  Research and Innovation Grant, 2021 
(Grant No.  DoRD/RIG/10-73/ 1592-A).  

\input{Dusty_Plasma_Revised.bbl}

\end{document}

%% file: Dusty_Plasma_Revised.bbl
%